\documentstyle[aps,preprint,epsf]{revtex}
\begin{document}
\draft

\title{Quantum chaos, statistical equilibrium and resonant radiative
capture of electrons by multicharged ions: Au$^{24+}$}

\author{G. F. Gribakin, A. A. Gribakina, and V. V. Flambaum}

\address{School of Physics, The University of New South Wales,
Sydney 2052, Australia}

\date{\today}

\maketitle

\tightenlines

\begin{abstract}

We show that the spectrum and eigenstates of open-shell multicharged
atomic ions near the ionization threshold are chaotic, as a result of
extremely high level densities of multiply excited electron states
($10^3$ eV$^{-1}$ in Au$^{24+}$) and a strong configuration mixing. This
complexity enables one to use statistical methods to analyze the system. The
orbital occupation numbers obey the Fermi-Dirac distribution, and temperature
can be introduced. We show that radiative capture of electrons through the
multielectron resonant states is strongly enhanced compared to the direct
radiative recombination.

\end{abstract}

\pacs{PACS numbers: 31.50.+w, 34.80.Lx, 32.70.Cs, 05.30.Fk}

%****************************************************************************

In this paper we investigate the spectrum and eigenstates of a multicharged
positive ion at energies close to its ionization threshold $I$. Using
Au$^{24+}$ ($I=750$ eV) as an example, we show that this spectrum is
dominated by multiple electron excitations into a few low-lying unoccupied
orbitals. As a result, it is extremely dense, with level spacings
$\sim 1$ meV between the states of a given total angular momentum and parity
$J^\pi $. The electron Coulomb interaction induces a strong mixing of the
multiply-excited configurations, which leads to a statistical equilibrium
in the system. The latter is similar to a thermal equilibrium, and
variables such as temperature can be introduced to describe it. This enables
one to use statistical methods in the situation where a full dynamical
quantum calculation is simply impossible because of the enormous size of the
Hilbert space ($\gtrsim 10^5$ for Au$^{24+}$).

We apply this approach to the problem of radiative capture of low-energy
electrons by multicharged positive ions, and show that the contribution of
resonant {\em multielectronic} recombination, which proceeds via electron
capture into the multiply-excited compound states, is much greater than that
of the direct radiative recombination. Our calculation gives a quantitative
explanation of huge enhancements of the recombination rates, and removes the
``enormous discrepancies between theoretical and experimental rate
coefficients'' \cite{Hoffknecht:98}. The situation here turns out to be
similar to the radiative neutron capture by complex nuclei [$(n,\gamma )$
reaction] where the resonance mechanism involving the compound nucleus states
is also much stronger than the direct capture \cite{Flambaum:84}.

So far the enhanced recombination at low electron energies $\lesssim 1$ eV has
been observed for a number of ions \cite{note}. Its magnitude ranges from a
factor of about ten for Ar$^{13+}$, Au$^{50+}$, Pb$^{53+}$, and U$^{28+}$
\cite{Gao:95}, to over a hundred for Au$^{25+}$ \cite{Hoffknecht:98}.
This enhancement is sensitive to
the electronic structure of the target, e.g., the recombination rates of
Au$^{49+}$ and Au$^{51+}$ are much smaller than that of Au$^{50+}$
\cite{Gao:95}. For few-electron ions, e.g., C$^{4+}$, Ne$^{7+}$ and
Ar$^{15+}$ \cite{Schennach:94,Zong:97,Schuch:97}, the observed rates are
described well by the sum of the direct and dielectronic recombination rates.
In more complicated cases, like U$^{28+}$ or Au$^{25+}$, the questions of what
are the particular resonances just above the threshold and how they contribute
to the recombination ``remain a mystery'' \cite{Mitnik:98}.

\begin{figure}[t]
\vspace{8pt}
\hspace{-35pt}
\epsfxsize=11cm
\centering\leavevmode\epsfbox{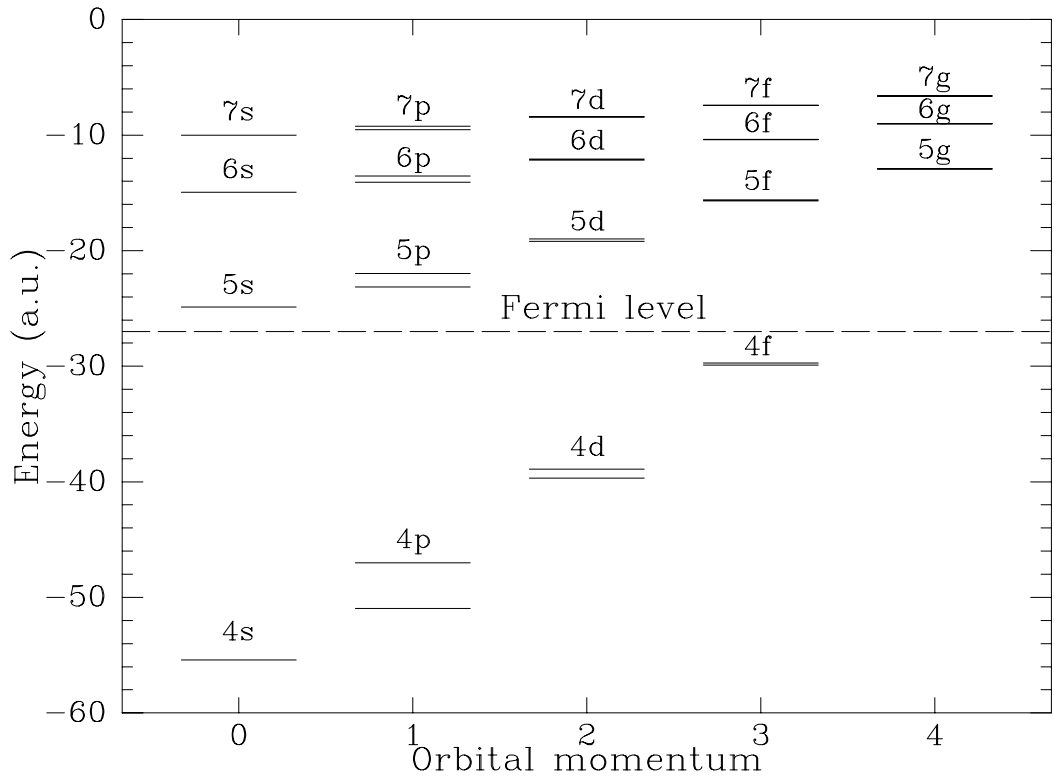}
\vspace{8pt}
\caption{Electron orbitals of Au$^{24+}$ from the Dirac-Fock
calculation of its ground state $1s^2\dots 4f^9$.}
\label{fig:orb}
\end{figure}

Let us consider the problem of electron recombination on Au$^{25+}$. Due to
electron correlations the electron can be captured into an excited
state of the compound Au$^{24+}$ ion. Au$^{24+}$ has 55 electrons. Its ground
state belongs to the $1s^2\dots 4f^9$ configuration. Figure \ref{fig:orb}
shows the energies of its relativistic orbitals $nlj$ from a self-consistent
Dirac-Fock calculation. The energy of the highest orbital occupied in the 
ground state is $\varepsilon _{4f_{7/2}}=-29.7$ a.u. Our relativistic
configuration-interaction (CI) calculation of the ground states of
Au$^{24+}4f^9$ and Au$^{24+}4f^8$ shows that they are characterized by
$J=\frac{15}{2}$ and 6, and their total energies are $-18792.36$ and
$-18764.80$ a.u., respectively. Thus, the ionization threshold of Au$^{24+}$
is $I=27.56$ a.u.$=750$ eV.

The excited states of the ion are generated by transferring electrons
from the ground state into the unoccupied orbitals above the Fermi level, or
into the partially filled $4f$ orbital. We are  interested in the excitation
spectrum of Au$^{24+}$ near its ionization threshold. This energy (27.5 a.u.)
is sufficient to push up a few of the $4f$ electrons, and even excite
one or two electrons from the $4d$ orbital \cite{note1}. Thus, we consider
Au$^{24+}$ as a system of $n=19$ electrons above the inactive Kr-like
$1s^2\dots 4p^6$ core.

The number of multielectron states obtained by distributing 19 electrons over
31 relativistic orbitals, $4d_{3/2}$ through to $7g_{9/2}$, is enormous, even
if we are only interested in the excitation energies below 27.5 a.u. It is
impossible to perform any CI calculation for them. However, there is another
simpler way to analyze the spectrum. The scale of the configuration interaction
strength is determined by the two-body Coulomb matrix elements. Their typical
size in Au$^{24+}$ is about 1 a.u., i.e., much smaller than $I$. Configuration
mixing aside, the CI does not shift the mean energies of the configurations.
Therefore, we can construct the excitation spectrum of Au$^{24+}$ by
calculating the mean energies $E_i$ of the configurations,
\begin{equation}\label{eq:Ek}
E_i=E_{\rm core}+\sum _a\epsilon _an_a+
\sum _{a\leq b}\frac{n_a(n_b-\delta _{ab})} {1+\delta _{ab}}U_{ab},
\end{equation}
and the numbers of many-electron states in each of them,
$N_i=\prod _a g_a!/[n_a!(g_a-n_a)!]$, where $n_a$ are the orbital occupation
numbers in the configuration ($\sum _a n_a=n$), $\epsilon _a$ is the
single-particle energy of the orbital $a$ in the field of the core,
$g_a=2j_a+1$, and $U_{ab}$ is the average Coulomb interaction between the
electrons in orbitals $a$ and $b$ (direct $-$ exchange).

\begin{figure}[t]
\vspace{8pt}
\hspace{-35pt}
\epsfxsize=12.0cm
\centering\leavevmode\epsfbox{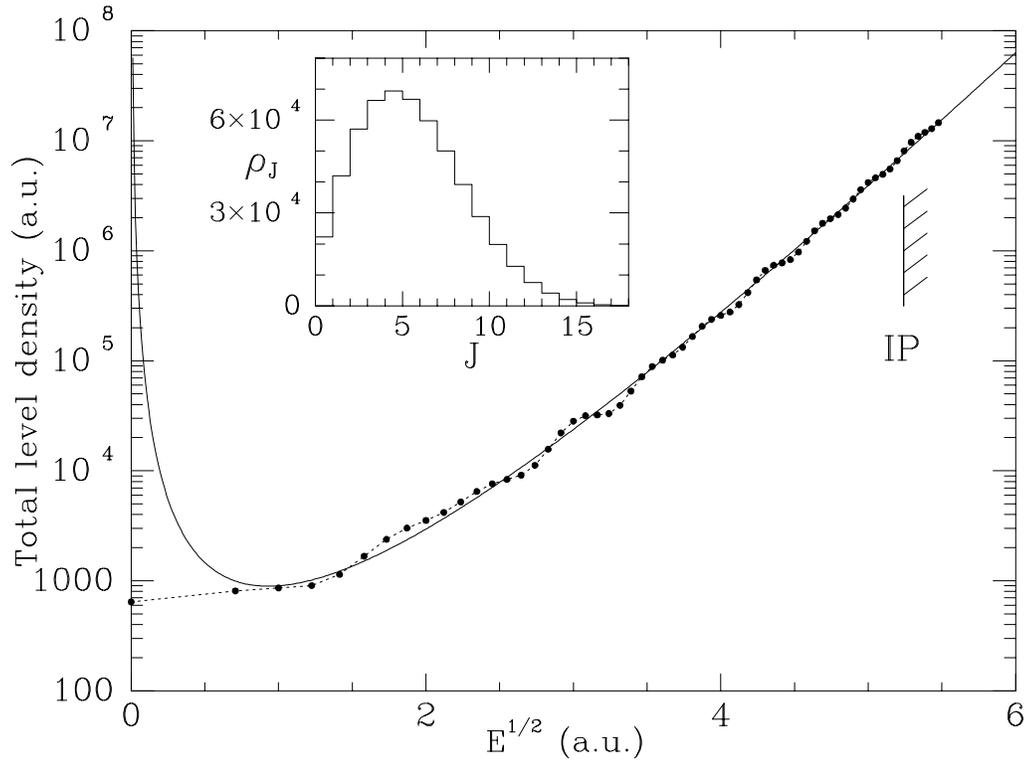}
\vspace{8pt}
\caption{Level density in Au$^{24+}$. Full circles --  numerical calculation.
Solid line -- analytical fit, Eq. (\ref{eq:fit}). The inset shows level
densities at $E\approx I$ for different $J$.}
\label{fig:dens}
\end{figure}

We find that there are 9000 configurations within 35 a.u. of the Au$^{24+}$
ground state. They comprise a total of $2.1\times 10^8$ many-electron states,
which corresponds to about $5\times 10^5$ states in each $J^\pi $ manifold.
Figure \ref{fig:dens} shows the total density of states
\begin{equation}\label{eq:rho}
\rho (E)=\sum _iN_i\delta (E-E_i)
\end{equation}
averaged over 1 a.u. energy intervals, as a function of $\sqrt{E}$, where $E$
is the excitation energy above the ground state. The inset presents a
break-up of the total density near the ionization threshold into the densities
of states with given $J$: $\rho (E)=\sum _J(2J+1)\rho _J(E)$. The most
abundant values are $J=\frac{5}{2}\,$--$\,\frac{15}{2}$. For a given parity
the density of such states at $E\approx I$ is
$\rho _{J^\pi }\approx 3.5\times 10^4$ a.u., which corresponds to a mean level
spacing $D=1/\rho _{J^\pi }\sim 1$ meV. Figure \ref{fig:dens} demonstrates the
characteristic $\rho \propto \exp (a\sqrt{E})$ behaviour of the level density
predicted by the Fermi-gas model \cite{Bohr:69}, where $a$ is related to the
single-particle level density at the Fermi level
$g(\varepsilon _F)=3a^2/2\pi^2$. A Fermi-gas model ansatz
\begin{equation}\label{eq:fit}
\rho (E)=AE^{-\nu }\exp (a\sqrt{E})~,
\end{equation}
with  $A=31.6$, $\nu =1.56$, and $a=3.35$ gives an accurate fit of the level
density at $E>1$ a.u. The corresponding $g(\varepsilon _F)=1.7$ a.u. is in
agreement with the orbital spectrum in Fig. \ref{fig:orb}. For most abundant
$J ^\pi$ states $\rho _{J^\pi }(E)$ is given by Eq. (\ref{eq:fit}) with
$A_{J^\pi }\approx 0.15$.

At first sight the huge level density makes the Au$^{24+}$ problem very
complicated. In reality this complexity enables one to use statistical
methods to describe the system. The interaction between multiply-excited
configuration states mixes them completely, and they
loose their individual features. In this regime the spectral statistics
are close to those of a random matrix ensemble, the eigenstates cannot be
characterized by any quantum numbers except the exact ones (energy and
$J^\pi $), and the orbital occupation numbers deviate prominently from
integers. This regime can be described as many-body quantum chaos. We have
extensively studied it in direct numerical calculations for the rare-earth
atom of Ce -- a system with four valence electrons
\cite{Flambaum:94,Flambaum:96,Flambaum:98}.

\begin{figure}[t]
\vspace{8pt}
\hspace{-35pt}
\epsfxsize=15.0cm
\epsfysize=12cm
%\centering\leavevmode\epsfbox{Aufig3.eps}
\epsfbox[-150 50 624 646]{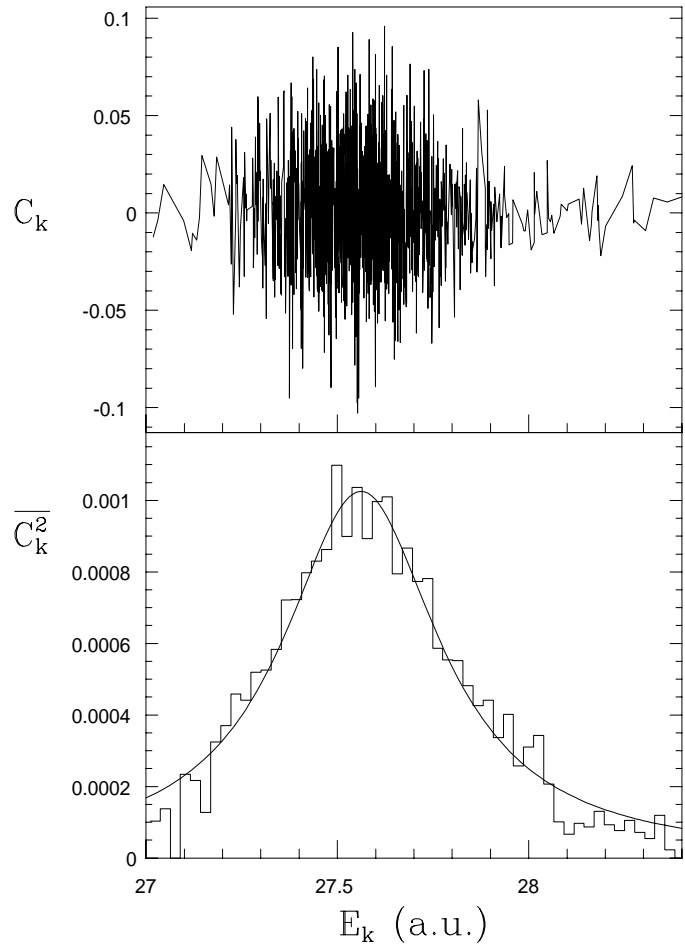}
\vspace{8pt}
\caption{Components of a 590th $J^\pi =\frac{13}{2}^+$ eigenstate from a
two-configuration calculation (top), and a fit of
$\protect \overline{C_k^2}(E)$ by the Breit-Wigner formula (bottom).}
\label{fig:comp}
\end{figure}

The strength of the configuration mixing is characterized by the spreading
width $\Gamma _{\rm spr}$. For a configuration basis state $\Phi _k$ with
energy $E_k$ it defines the energy range $|E-E_k|\lesssim \Gamma _{\rm spr}$
of eigenstates in which this basis state noticeably participates. By the same
token it shows
that a particular eigenstate $\Psi =\sum _kC_k\Phi _k$ contains a large number
$N\sim \Gamma _{\rm spr}/D$ of {\em principal components} -- basis states
characterized by $C_k\sim 1/\sqrt{N}$. Outside the spreading width
$C_k$ decrease. Apart from this, $C_k$ behave like random variables
\cite{Flambaum:94}. The effect of spreading is approximated well by
the Breit-Wigner shape \cite{Bohr:69}
\begin{equation}\label{eq:BW}
\overline {C_k^2}(E)=N^{-1}\frac{\Gamma _{\rm spr}^2/4}
{(E_k-E)^2+\Gamma _{\rm spr}^2/4}~.
\end{equation}
The normalization $\sum _k\overline {C_k^2}=1$ yields $N=\pi\Gamma /2D$.
In systems with small level spacings $D$ the number of principal components
$N$ can be very large. It reaches several hundreds in Ce, and $10^6$ in
complex nuclei. In Fig. \ref{fig:comp} we illustrate this behaviour by the
results of a CI calculation for two Au$^{24+}$ configurations near the
ionization threshold: $4f_{5/2}^34f_{7/2}^35p_{1/2}5p_{3/2}5f_{7/2}$ and
$4f_{5/2}^34f_{7/2}^35p_{1/2}5d_{3/2}5g_{7/2}$. These two configurations
produce 1254 $J^\pi =\frac{13}{2}^+$ states. Their mixing is complete, since
the weight of each configuration in every eigenstate is about 50\%.
A Breit-Wigner fit of the mean-squared components yields $N=975$ and
$\Gamma _{\rm spr}=0.50$ a.u. The spreading width is related to the
mean-squared off-diagonal Hamiltonian matrix element and the mean level
spacing as $\Gamma _{\rm spr}\simeq 2\pi \overline{H_{ij}^2}/D$ \cite{Bohr:69}.
When more configurations are included, both $D$ and $\overline{H_{ij}^2}$
decrease, and $\Gamma _{\rm spr}$ does not change much. If one were able to do
a full-scale CI calculation near the ionization threshold of Au$^{24+}$
eigenstates with $N=(\pi /2)\Gamma _{\rm spr} \rho _{J^\pi}
\sim 3\times 10^4$ would be obtained.

\begin{figure}[t]
\hspace{-35pt}
\epsfysize=13cm
\epsfbox[-150 0 624 646]{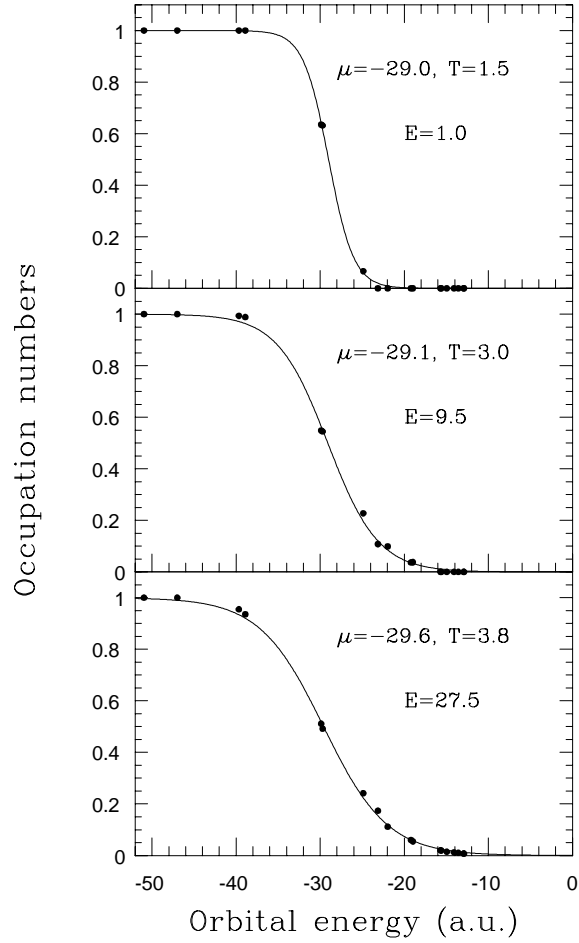}
%\centering\leavevmode\epsfbox{Aufig4.eps}
\vspace{-35pt}
\caption{Orbital occupation numbers in Au$^{24+}$ calculated at $E=1$, $9.5$,
and $27.5$ a.u. (circles), compared with the Fermi-Dirac distribution
(solid line).}
\label{fig:FD}
\end{figure}

\begin{figure}[t]
\vspace{8pt}
\hspace{-55pt}
\epsfxsize=7cm
\centering\leavevmode\epsfbox{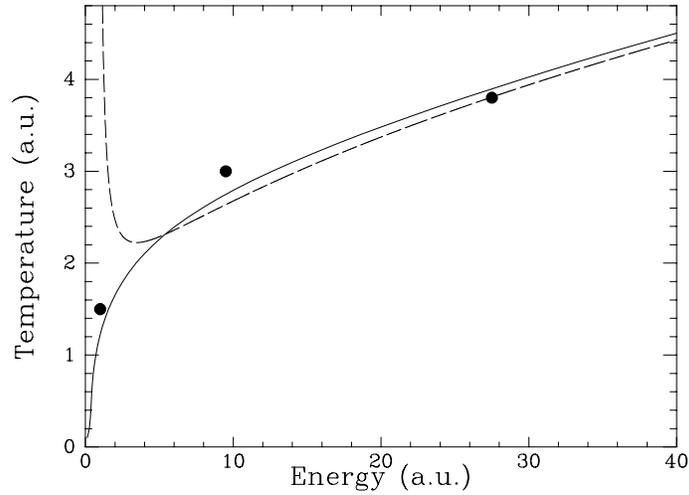}
\vspace{8pt}
\caption{Temperature vs energy for Au$^{24+}$. Solid line - canonical
definition; dashed line - statistical definition;
solid circles - Fermi-Dirac fits of the occupation numbers.}
\label{fig:temp}
\end{figure}

The spreading of the basis states due to configuration interaction introduces
a natural statistical averaging in the system. Based on this averaging, a
statistical theory of finite Fermi systems of interacting particles can be
developed \cite{Flambaum:97}. It enables one to calculate various properties
of the system without diagonalizing the Hamiltonian, as sums over the basis
states, e.g., the mean occupations numbers,
$n_a(E)=\sum _k\overline{C_k^2}(E)n_a^{(k)}$, where $n_a^{(k)}$ is the
occupation number of the orbital $a$ in the basis state $k$. Using a simple
Gaussian model spreading we calculate mean orbital occupation numbers at
different excitation energies. Figure \ref{fig:FD} shows that the
result (circles) is described well by the Fermi-Dirac formula
$n_a=\{1+\exp [(\varepsilon _a-\mu )/T]\}^{-1}$. The temperature $T$ and
the chemical potential $\mu $ are chosen to give the best fit of the numerical
occupation numbers. On the other hand, the temperature can be
determined from the level density $\rho (E)$, Eq. (\ref{eq:rho}), through the
canonical average $E(T)=Z^{-1}\int e^{-E/T}E\rho (E)dE$,
where $Z=\int e^{-E/T}\rho (E)dE$, or from a statistical formula
$T^{-1}=d\ln [\rho (E)]/dE$, using the smooth fit (\ref{eq:fit}). The latter
yields $T\simeq 2\sqrt{E}/a$, characteristic of Fermi systems.
For energies above 3 a.u. all three definitions give close values,
Fig. \ref{fig:temp}. 

Let us now estimate the direct and resonant contributions to the recombination
rate of Au$^{25+}$. The direct radiative recombination cross section is
estimated by introducing an effective ionic charge $Z_i$ into the formula of
Bethe and Salpeter \cite{Bethe:57},
\begin{equation}\label{eq:sigmad}
\sigma ^{\rm (d)}=\frac{1.96\pi^2}{c^3}\,\frac{\rm Ryd}{\varepsilon }Z_i^2
\ln \left( \frac{Z_i}{n_0}\sqrt{\frac{\rm Ryd}{\varepsilon }}\right)~,
\end{equation}
where $\varepsilon $ is the electron energy and $n_0$ is the principal
quantum number of the lowest unoccupied ionic orbital (we use atomic units).
Using $Z_i=25$, $n_0=5$, and $\varepsilon =0.1$ eV we obtain
$\sigma ^{\rm (d)}\approx 7\times 10^{-17}$ cm$^{2}$, which gives a
rate of $\lambda =\sigma v= 1.3\times 10^{-9}$ cm$^3$s$^{-1}$, two orders of
magnitude smaller than the experimental
$\lambda =1.8\times 10^{-7}$ cm$^3$s$^{-1}$ at this energy
\cite{Hoffknecht:98}.

The energy-averaged cross section of the resonant radiative capture of a
low-energy $s$ electron is \cite{Landau:77}
\begin{equation}\label{eq:sigmar}
\sigma ^{\rm (r)}=\frac{\pi ^2}{\varepsilon }\,\frac{\Gamma _\gamma \Gamma _e}
{D(\Gamma _\gamma +\Gamma _e)}\approx \frac{\pi ^2}{\varepsilon }\,
\frac{\Gamma _\gamma}{D}~~~(\Gamma _e\gg \Gamma _\gamma ),
\end{equation}
where $\Gamma _\gamma $ and $\Gamma _e$ are the mean radiative and
autoionization (or elastic) widths of the resonances, $D$ is the mean
resonance spacing, and we drop the statistical weights of the initial and
intermediate ionic states. The relation $\Gamma _e\gg \Gamma _\gamma $ is
usually valid for a few lower partial waves, where the electron interaction
is stronger than the electromagnetic one.

The radiative width of the resonant state at energy $E\approx I$ is found by
summing the partial widths for all lower-lying states $E'=E-\omega $,
\begin{equation}\label{eq:width}
\Gamma _\gamma \approx \frac{3}{2J+1}\int _0^I\frac{4\omega ^3|d_\omega |^2}
{3c^3}\rho _{J^\pi }(I-\omega )d\omega ~,
\end{equation}
where the factor 3 accounts for $J'=J,~J\pm 1$, and $d_\omega $ is the reduced
dipole matrix element between the many-electron states. Because of the chaotic
structure of these states $d_\omega $ is suppressed compared to the typical
single-particle matrix element $d_0$: $d_\omega \sim d_0/\sqrt{N}$
\cite{Flambaum:84,Flambaum:94,Flambaum:96}. This estimate for systems with
dense chaotic spectra in fact follows from the dipole sum rule: the number of
lines in the spectrum is large, $\propto D^{-1}\propto N$, consequently, the
line strengths are small, $|d_\omega |^2\sim |d_0|^2N^{-1}$.

Using this estimate and calculating the integral in Eq. (\ref{eq:width})
by the saddle-point method we obtain
\begin{equation}\label{eq:sigres1}
\sigma ^{\rm (r)}=\frac{8\pi d_0^2}{(2J+1)c^3\varepsilon \Gamma _{\rm spr}}
\sqrt{\frac{2\pi}{3}}\rho _{J^\pi}(I-\omega _0)\omega _0^4~,
\end{equation}
where $\omega _0=6\sqrt{I}/a$ corresponds to the maximum of the the decay
photon spectrum in Eq. (\ref{eq:width}). This cross section has the same energy
dependence as $\sigma ^{\rm (d)}$. To evaluate its magnitude we use
$d_0\sim Z_i^{-1}$, $2J+1\approx 10$, and substitute $\Gamma _{\rm spr}=0.5$, 
$\omega _0=9.4$, and $\rho _{J^\pi}(I-\omega _0)=2.5\times 10^3$ a.u.
At $\varepsilon =0.1$ eV this gives $\sigma ^{\rm (r)}= 7\times 10^{-16}$
cm$^2$, therefore, $\sigma ^{\rm (r)}/\sigma ^{\rm (d)}= 10$, and we
obtain a factor of ten enhancement over the direct recombination.
It comes from the large effective number of final states in the radiative
width in Eq. (\ref{eq:width}) (numerically
$\Gamma _\gamma \sim 2\times 10^{-7}$ a.u.). If we include contributions of
the higher electron partial waves the enhancement will match the experimentally
observed values.

In summary, the resonant radiative capture mechanism fully explains the
strongly enhanced recombination rates observed for eV electrons on
multicharged ions. Its origin is in the high level densities of chaotic
multiply-excited electron states in multicharged ions. The size of the
enhancement is sensitive to the electron structure of the ion, which
determines the level density. We have shown that a statistical approach can be
applied to the analysis of this complex system. One can also use a statistical
theory to calculate mean-squared matrix elements between multiply
excited chaotic states in terms of single-particle amplitudes, occupation
numbers, $\Gamma _{\rm spr}$ and $D$
\cite{Flambaum:94,Flambaum:96,Flambaum:93}, and obtain accurate quantitative
information about the processes involving chaotic states and resonances.
At higher electron energies the resonant capture proceeds via so-called doorway
states \cite{Bohr:69} -- simple dielectronic autoionizing states, which are
then ``fragmented'' into the dense spectrum of multiply-excited resonances
(see \cite{Mitnik:98} and \cite{Flambaum:96} and Refs. therein).

\bibliographystyle{prsty}
% \bibliography{whole}

\begin{thebibliography}{10}
\bibitem{note1}Infinite Rydberg series belong to the single-particle aspect of
the problem, and we do not consider them here.
\end{thebibliography}

%*****************************************************************************
\figure

\end{document}